\documentstyle[12pt]{article}
\begin{document}
\pagestyle{empty} \setlength{\footskip}{2.0cm}
\setlength{\oddsidemargin}{0.5cm} \setlength{\evensidemargin}{0.5cm}
\renewcommand{\thepage}{-- \arabic{page} --}
\def\mib#1{\mbox{\boldmath $#1$}}
\def\bra#1{\langle #1 |}      \def\ket#1{|#1\rangle}
\def\vev#1{\langle #1\rangle} \def\dps{\displaystyle}
\newcommand{\bg}{B.
 Grz${{{\rm a}_{}}_{}}_{\hskip -0.18cm\varsigma}$dkowski}
\newcommand{\BG}{Bohdan
 GRZ${{{\rm A}_{}}_{}}_{\hskip -0.18cm\varsigma}$DKOWSKI}
   \def\thebibliography#1{\centerline{REFERENCES}
   \list{[\arabic{enumi}]}{\settowidth\labelwidth{[#1]}\leftmargin
   \labelwidth\advance\leftmargin\labelsep\usecounter{enumi}}
   \def\newblock{\hskip .11em plus .33em minus -.07em}\sloppy
   \clubpenalty4000\widowpenalty4000\sfcode`\.=1000\relax}\let
   \endthebibliography=\endlist
   \def\sec#1{\addtocounter{section}{1}\section*{\hspace*{-0.72cm}
     \normalsize\bf\arabic{section}.$\;$#1}\vspace*{-0.3cm}}
\vspace*{-1cm}\noindent
\hspace*{10.8cm}IFT-23-96\\
\hspace*{10.8cm}TOKUSHIMA 96-03\\
\hspace*{10.8cm}(hep-ph/9610306)\\

\vspace*{.5cm}

\begin{center}
\renewcommand{\thefootnote}{\dag}
{\large\bf Probing $\mib{C}\!\mib{P}$ Violation via Top Polarization
at NLC}$\,$\footnote{Talk presented by Z. Hioki at Circum-Pan-Pacific
Workshop on {\it High Energy Spin Physics'96}, October 2-4, 1996,
Kobe University, Kobe, Japan}
\end{center}

\vspace*{1.25cm}
\begin{center}
\renewcommand{\thefootnote}{*)}
{\sc Bohdan GRZ${{{\rm A}_{}}_{}}_{\hskip -0.18cm\varsigma}
$DKOWSKI$^{\:a),\:}$}\footnote{E-mail address:
\tt bohdan.grzadkowski@fuw.edu.pl}
and
\renewcommand{\thefootnote}{**)}
{\sc Zenr\=o HIOKI$^{\:b),\:}$}\footnote{E-mail address:
\tt hioki@ias.tokushima-u.ac.jp}
\end{center}

\vspace*{1.2cm}
\centerline{\sl $a)$ Institute for Theoretical Physics,\ Warsaw 
University}
\centerline{\sl Ho\.za 69, PL-00-681 Warsaw, POLAND} 

\vskip 0.3cm
\centerline{\sl $b)$ Institute of Theoretical Physics,\ 
University of Tokushima}
\centerline{\sl Tokushima 770, JAPAN}

\vspace*{1.8cm}
\centerline{ABSTRACT}

\vspace*{0.4cm}
\baselineskip=20pt plus 0.1pt minus 0.1pt
Possible $C\!P$-violation in top-quark couplings is discussed. It is
shown that the lepton-energy distributions in $e^+ e^-\to t\bar{t}\to
\ell^+\ell^-X\,/\,\ell^\pm X$ at next linear colliders (NLC) could
give us useful information for this study. The statistical
significance of $C\!P$-violating-parameter determination is
estimated.
\vfill
\newpage
\renewcommand{\thefootnote}{\sharp\arabic{footnote}}
\pagestyle{plain} \setcounter{footnote}{0}
\baselineskip=21.0pt plus 0.2pt minus 0.1pt

\sec{Introduction}
By the discovery of the top quark, the fermion spectrum required by
the standard EW theory (SM) has been completed. Still, it is to be
seen if the third generation is a copy of the first and second ones
or any new interactions exist in top-quark couplings. Studying $C\!P$
violation via top-quark polarization is an interesting approach to
this problem. This is because $(i)$ the $C\!P$ violation in the
top-quark couplings induced within the SM is far negligible and
$(ii)$ a lot of information on the top quark is to be transferred to
the secondary leptons without getting obscured by the hadronization
effects thanks to huge $m_t$.

Here we would like to show our recent work on this topic \cite{BGZH}.
First we describe why $t\bar{t}$ productions are useful for our study
in a little more detail. Next we show how the signal of $C\!P$
non-conservation that might occur there would be affected by another
possible $C\!P$-violating interaction in the decay process. We then
introduce a new asymmetry to catch their combined signal effectively,
and discuss the expected statistical uncertainties in the
corresponding-parameter determination.

\sec{Polarized Top Production and $\mib{C}\!\mib{P}$ Violation}
Since $t\bar{t}$ pairs are produced mainly through the vector-boson
exchange, the handedness of $t$ and $\bar{t}$ must be the same.
Consequently, the helicities of $t\bar{t}$ would be $(+-)$ or $(-+)$
if the top mass were much smaller than $\sqrt{s}$. However, since the
observed $m_t$ is by no means negligible, we will also face copious
production of $(++)$ and $(--)$ states. For example, $\sigma_{tot}
(e^+e^-\!\to t\bar{t})$ is estimated to be 0.60 pb for $\sqrt{s}=$500
GeV (and $m_t=$180 GeV) within the SM, in which $N(-+):N(+-):N(--):
N(++)$ is $4.8:3.4:0.9:0.9$, where $N(\cdots)$ denotes the number of
the $t\bar{t}$ pairs with indicated helicities.

We can take advantage of this fact to explore $C\!P$ properties of
the $t\bar{t}$ state. That is, $|\!-\!-\rangle$ and $|\!+\!+\rangle$
are transformed into each other by $C\!P$ operation, and consequently
the difference between $N(--)$ and $N(++)$ could be a useful measure
of $C\!P$ violation. Although what we can observe in experiments are
not the top quarks but products of their subsequent decays, the
energy spectrum of $\ell^+$ and $\ell^-$ in $e^+e^-\!\to t\bar{t} \to
\ell^+\ell^- X/\ell^\pm X$ can be a good measure of $N(--)-N(++)$, as
we will see.

The leptonic energy spectrum has been studied in the existing 
literature \cite{CKPAS}.\footnote{We did not aim to give a complete
   reference list owing to limited space. See the lists of our
   articles \cite{BGZH}.}\ 
However, $C\!P$-violating interactions were assumed only in the
$t\bar{t}\gamma/Z$ vertices and the standard-model vertex was used
for the $t\to bW$ decay in those articles. In order to perform a
consistent analysis of the top-quark couplings, we computed the
spectrum assuming that both the $t\bar{t}\gamma/Z$ vertices and the
$tbW$ vertex include non-standard $C\!P$-violating form factors.

\sec{Effects of Non-Standard Top Decay}
As already mentioned, $\delta\equiv[\,N(--)-N(++)\,]/N(all)$ is a
good measure of $C\!P$ violation in $t\bar{t}$ productions. If there
is no $C\!P$ violation in the $tbW$ vertex, the energy-spectrum
asymmetry $a(x)$ defined as
\begin{equation}
a(x)\equiv\frac{d\sigma^-/dx-d\sigma^+/dx}{d\sigma^-/dx+d\sigma^+/dx}
\label{asy}
\end{equation}
is known to be proportional to $\delta$ \cite{CKPAS}, where $d
\sigma^\pm/dx$ are the $\ell^\pm$ distributions in the reduced lepton
energy $x\equiv 2E\sqrt{(1-\beta)/(1+\beta)}/m_t$ with $E$ being the
energy of $\ell^\pm$ in the $e^+e^-$ c.m. system and $\beta\equiv
\sqrt{1-4m_t^2/s}$. In such a case, $a(x)$ would serve as a useful
observable to measure $C\!P$ violation.

However, if $C\!P$ non-conservation occurs also in the $tbW$ vertex,
it becomes
$$
a(x)=\frac{-2(\delta/\beta)\,g(x) +\hbox{Re}{(f_2^R-\bar{f}_2^L)}
[\,\delta\! f(x)+\eta\: \delta g(x)\,]}{2\,[\,f(x)+\eta\: g(x)\,]},
$$
where the functions $f(x)$, $g(x)$, $\delta\!f(x)$ and $\delta g(x)$,
and the parameter $\eta$ are defined in \cite{BGZH}. The top
couplings with $\gamma/Z$ and $W$ we assumed for computing the above
$a(x)$ are
\begin{equation}
{\mit\Gamma}^\mu=\frac{g}{2}\:\bar{u}(p_t)
\biggl[\,\gamma^\mu(A_v-B_v\gamma_5)
+\frac{(p_t-p_{\bar{t}})^\mu}{2m_t}(C_v-D_v\gamma_5)\,\biggr]v(p_t),
\label{vtt}
\end{equation}
\begin{eqnarray}
&&{\mit\Gamma}^{\mu}=-{g\over\sqrt{2}}V_{tb}\:
\bar{u}(p_b)\biggl[\,\gamma^{\mu}(f_1^L P_L +f_1^R P_R)
-{{i\sigma^{\mu\nu}k_{\nu}}\over M_W}
(f_2^L P_L +f_2^R P_R)\,\biggr]u(p_t),\ \ \ \ \\
&&\bar{\mit\Gamma}^{\mu}=-{g\over\sqrt{2}}V_{tb}^*\:
\bar{v}(p_t)\biggl[\,\gamma^{\mu}(\bar{f}_1^L P_L +\bar{f}_1^R P_R)
-{{i\sigma^{\mu\nu}k_{\nu}}\over M_W}
(\bar{f}_2^L P_L +\bar{f}_2^R P_R)\,\biggr]v(p_b),
\end{eqnarray}
where $g$ is the SU(2) gauge-coupling constant, $v=\gamma/Z$,
$P_{L/R}\equiv(1\mp\gamma_5)/2$, $V_{tb}$ is the $(tb)$ element of
the Kobayashi-Maskawa matrix and $k$ is $W$'s momentum. Measuring
asymmetries like $a(x)$ under such a circumstance is a challenging
task since it is differential and therefore the expected statistics
cannot be high. That is why we looked for another more effective
quantity.

\sec{$\mib{C}\!\mib{P}$-Violating Asymmetry}
We introduce the following $C\!P$-violating asymmetry:
\begin{equation}
A_{\ell\ell}\equiv
\Bigl(\int\!\!\!{\int_{}}_{x<\bar{x}}\!\!\!\!\!
dxd\bar{x}\frac{d^2\sigma}{dxd\bar{x}}
-\int\!\!\!{\int_{}}_{x>\bar{x}}\!\!\!\!\!
dxd\bar{x}\frac{d^2\sigma}{dxd\bar{x}}\Bigr)
\Bigl/\int\!\!\!\int dxd\bar{x}\frac{d^2\sigma}{dxd\bar{x}}\:,
\label{asymm}
\end{equation}
where $x$ and $\bar{x}$ are the reduced energies of $\ell^+$ and
$\ell^-$ respectively.

For the SM parameters $\sin^2\theta_W=0.2325$, $M_W=80.26$ GeV, $M_Z=
91.1884$ GeV, ${\mit\Gamma}_Z=2.4963$ GeV and $m_t=180$ GeV, it
becomes
\begin{eqnarray}
&&A_{\ell\ell}
=0.3089\:{\rm Re}(f^R_2-\bar{f}^L_2)+0.3638\:{\rm Re}(D_\gamma)
  +0.0609\:{\rm Re}(D_Z)  \nonumber \\
&&\phantom{A_{\ell\ell}}
=0.3089\:{\rm Re}(f^R_2-\bar{f}^L_2)-0.3441\:\xi,
\end{eqnarray}
where $\xi$ is given by $\xi=-\delta/\beta$ and therefore
characterizes the $C\!P$ violation in the production process. For
${\rm Re}(f^R_2)=-{\rm Re}(\bar{f}^L_2)={\rm Re}(D_\gamma)={\rm Re}
(D_Z)=0.2$, e.g., we have $A_{\ell\ell}=0.2085$. Its statistical
error for $N_{\ell\ell}$ events is estimated thereby to be
$$
{\mit\Delta}A_{\ell\ell}=\sqrt{(1-A_{\ell\ell}^2)/N_{\ell\ell}}
=0.9780/\sqrt{N_{\ell\ell}}.
$$

Since $\sigma_{e\bar{e}\to t\bar{t}}=0.60$ pb for $\sqrt{s}=500$
GeV, the expected number of events is $N_{\ell\ell}=600\,
\epsilon_{\ell\ell}L B_\ell^2$, where $\epsilon_{\ell\ell}$ stands
for the $\ell^+\ell^-$ tagging efficiency ($=\epsilon_\ell^2\:$;
$\epsilon_{\ell}$ is the single-lepton-detection efficiency), $L$ is
the integrated luminosity in fb$^{-1}$ unit, and $B_\ell(\simeq
0.22)$ is the leptonic branching ratio for $t$. Consequently we
obtain ${\mit\Delta}A_{\ell\ell}=0.1815/\sqrt{\epsilon_{\ell\ell}L}$,
and thereby we are able to compute the statistical significance of
the asymmetry observation $N_{S\!D}=|A_{\ell\ell}|/{\mit\Delta}
A_{\ell\ell}$. For $L=50$ fb$^{-1}$ and $\epsilon_{\ell\ell}=0.5$,
for example, we get $N_{S\!D}=5.7$. This means we can confirm
$A_{\ell\ell}$ to be non-zero at $5.7\sigma$ level concerning the
statistical uncertainty.

\sec{Precision of Parameter Measurements}
By using $A_{\ell\ell}$, we will be able to observe a combined signal
of $C\!P$ violation in the productions and decays. In order to
study the new interactions in more detail, however, it is
indispensable to separate the parameters in the production and the
decay, i.e., $\xi$ and $\hbox{Re}{(f^R_2-\bar{f}^L_2)}$. For this
purpose we applied the optimal procedure of Ref.\cite{opt-96} both to
the $\ell^+\ell^-$ distribution and the $\ell^\pm$ distributions.

Here we show only the numerical results.\\
Through the double distribution, we can determine the parameters with
the following statistical uncertainties:
\begin{equation}
{\mit\Delta}{\rm Re}(f^R_2-\bar{f}^L_2)
=8.3824/\sqrt{N_{\ell\ell}},\ \ \
{\mit\Delta}\xi=7.1651/\sqrt{N_{\ell\ell}}\,. 
\label{D-delta}
\end{equation}
On the other hand, we get ${\mit\Delta}{\rm Re}(f^R_2)=7.1136/
\sqrt{N_{\ell}}$ and ${\mit\Delta}\xi=12.3285/\sqrt{N_{\ell}}$ from
the $\ell^+$ distribution, and analogous for ${\mit\Delta}{\rm Re}
(\bar{f}^L_2)$ and ${\mit\Delta}\xi$ from the $\ell^-$ distribution.
Since these two distributions are statistically independent, we can
combine them as
\begin{equation}
{\mit\Delta}{\rm Re}(f^R_2-\bar{f}^L_2)
=10.0601/\sqrt{N_\ell},\ \ \ {\mit\Delta}\xi=8.7176/\sqrt{N_\ell}\,. 
\label{S-delta}
\end{equation}

It is premature to conclude from Eqs.(\ref{D-delta}) and
(\ref{S-delta}) that we get a better precision in the analysis with
the double distribution. As it could be observed in the numerators in
Eqs.(\ref{D-delta}, \ref{S-delta}), {\it we lose some information
when integrating the double distribution on one variable}. However,
{\it the size of the expected uncertainty depends also on the number
of events}. That is, $N_{\ell\ell}$ is suppressed by the extra factor
$\epsilon_\ell B_{\ell}$ comparing to $N_{\ell}$. This suppression is
crucial even if we could achieve $\epsilon_\ell =1$. For $N$ pairs of
$t\bar{t}$ and $\epsilon_\ell=1$ we obtain
$$
{\mit\Delta}{\rm Re}(f^R_2-\bar{f}^L_2)=38.1018/\sqrt{N},\ \ \
{\mit\Delta}\xi=32.5686/\sqrt{N}
$$
from the double distribution, while
$$
{\mit\Delta}{\rm Re}(f^R_2-\bar{f}^L_2)=21.4484/\sqrt{N},\ \ \
{\mit\Delta}\xi=18.5859/\sqrt{N}
$$
from the single distribution. Therefore we may say that the
single-lepton distribution analysis is more advantageous for
measuring the parameters individually.

\sec{Summary}
Next-generation linear colliders of $e^+ e^-$ will provide a cleanest
environment for studying top-quark interactions. There, we shall be
able to perform detailed tests of the top-quark couplings to the
vector bosons and either confirm the SM simple generation-repetition
pattern or discover some non-standard interactions.

What we discussed here are the non-standard $C\!P$-violating
interactions in the $t\bar{t}$ productions and their subsequent
decays. If the top-quark decay was described by the SM interactions,
then we would have a useful compact formula for a measurement of
$C\!P$ violation in the $t\bar{t}\gamma/Z$ vertices via the
final-lepton-energy asymmetry (\ref{asy}). However, in general,
$C\!P$ violation may also enter through the top-decay process at the
same strength as it does for the production. Therefore, we have
assumed the most general $C\!P$-violating interactions both in the
production and in the decay vertices in order to perform a consistent
analysis.

We introduced a new asymmetry $A_{\ell\ell}$ in Eq.(\ref{asymm}),
which was shown to work quite effectively. Then, applying the optimal
procedure \cite{opt-96}, we studied the statistical significances of
$C\!P$-violation-parameter determination in analyses with the
double- and single-lepton energy distributions. Taking into account
the size of the leptonic branching ratio of the top quark and its
detection efficiency, we conclude that the use of the single-lepton
distribution is more advantageous to determine each $C\!P$-violation
parameter separately.

\vspace*{0.7cm}
\centerline{ACKNOWLEDGMENTS}

\vspace*{0.3cm}
One of us (Z.H.) is grateful to T. Morii and the organizing committee
for their warm hospitality. This work is supported in part by the
Committee for Scientific Research (Poland) under grant 2\ P03B 180
09, by Maria Sk\l odowska-Curie Joint Found II (Poland-USA) under
grant MEN/NSF-96-252, and by the Grant-in-Aid for Scientific Research
No.06640401 from the Ministry of Education, Science, Sports and
Culture (Japan).

\vspace*{0.7cm}

\end{document}